\documentclass[preprint2]{emulateapj-rtx4}

\newcommand{\kms}{{\km\second^{-1}}}
\newcommand{\solarmass}{{\,{\textnormal{M}}_\odot}}
\newcommand{\AU}{{\,\textnormal{AU}}}
\newcommand{\km}{\,{\textnormal{km}}}
\newcommand{\second}{\,{\textnormal{s}}}
\newcommand{\stellarmass}{{{\textnormal{M}}_*}}

\newcommand{\dechms}[4]{$#1^{\rm h}#2^{\rm m}#3\mbox{$^{\rm s}\mskip-7.6mu.\,$}#4$}

\newcommand{\decdms}[4]{$+#1^{\circ}#2'#3\mbox{$''\mskip-7.6mu.\,$}#4$}

\slugcomment{To appear in ApJ}

\shorttitle{A rotating molecular jet in Perseus}
\shortauthors{Pech et al.}

\begin{document}


\title{A rotating molecular jet from a Perseus protostar}


\author{Gerardo Pech, Luis A. Zapata, Laurent Loinard\footnote{Max-Planck-Institut f\"{u}r 
Radioastronomie, Auf dem H\"{u}gel 69  53121 Bonn, Alemania}, and Luis F. Rodr\'\i guez}
\affil{Centro de Radioastronom{\'\i}a y Astrof{\'\i}sica, Universidad 
Nacional Aut\'onoma de M\'exico, Morelia 58090, Mexico}


\begin{abstract}
We present $^{12}$CO(2-1) line and 1.4 mm continuum archival observations, made with the Submillimeter 
Array, of the outflow HH 797 located in the IC 348 cluster in Perseus.  The continuum emission
is associated with a circumstellar disk surrounding the class 0 object IC 348-MMS/SMM2, a very young solar 
analog. The line emission, on the other hand, delineates a collimated outflow, and reveals velocity asymmetries 
about the flow axis over the entire length of the flow. The amplitude of  velocity differences is of order 2 km 
s$^{-1}$ over distances of about 1000 AU, and we interpret them as evidence for jet rotation --although we also 
discuss alternative possibilities. A comparison with theoretical models suggests that the magnetic field lines 
threading the protostellar jet might be anchored to the disk of a radius of  about 20 AU. 
 \end{abstract}





\section{Introduction}

Molecular outflows are one of the first manifestations of the formation of a new star.  
They are thought to play an essential role in the removal of angular momentum from 
collapsing dense cores, which eventually enable them to contract to stellar sizes.  
Although the details remain unclear, the general consensus is that outflows are driven 
by rotating magnetic fields anchored to the disk-star system \citep{pu2007,shang2007}.  
Accretion disks are known to rotate, so the material ejected from them should inherit
a toroidal angular momentum component \citep{fe2011}.  Indeed, observational evidence 
for this component have been presented for several outflows driven by young stellar 
objects \citep{cho2011,zap2010,Lau2009,lee2009,lee2008}.

Arguably the most direct evidence has been presented by Launhardt et al. (2009) 
who reported millimeter interferometric observations of the isolated Bok globule 
CB 26.  Their $^{12}$CO(2-1) line observations revealed  the presence of a 
systematic velocity gradient perpendicular to the flow axis along its entire length.  
Zapata et al (2010), on the other hand, reported three independent $^{12}$CO(2-1) 
and SO(6-5) observations toward the Ori-S6 molecular outflow.  All three observations 
revealed velocity asymmetries about the outflow axis  which are suggestive of rotation 
at different size scales.

IC 348 is a young star cluster ($\sim$ 2 Myr) located near the eastern edge of the 
Perseus dark cloud complex at about 300 pc from the sun (Muench et al. 2007; 
Luhman et al. 2003).  It contains more than a dozen protostars, many of them driving
outflows. The IC 348-MMS/SMM2 source is a dense condensation within that region 
which drives a large collimated north-south CO ouflow \citep{tafa2006,hat2009}.  The 
optical and infrared counterpart  of this molecular outflow is the Herbig-Haro system 
HH 797 \citep{mac1994,eis2003,wal2005,wal2006}.  The low radial velocities 
displayed  by the outflow (blue to the north and red to the south) and its large extension 
($\sim$ 0.5 pc) suggest that it is nearly in the plane of the sky.

In this study, we present millimeter observations, made with the Submillimeter 
Array\footnote{The Submillimeter Array (SMA) is a joint project between the Smithsonian
Astrophysical Observatory and the Academia Sinica Institute of Astronomy and 
Astrophysics, and is funded by the Smithsonian Institution and the Academia Sinica.}
(SMA), of the IC 348-MMS/SMM2 object and its associated HH 797 outflow. These
interferometric observations reveal velocity asymmetries about the flow axis over
the entire length of the flow, that will be interpreted as jet rotation.

\begin{figure}[ht]
\begin{center}
\includegraphics[scale=0.33]{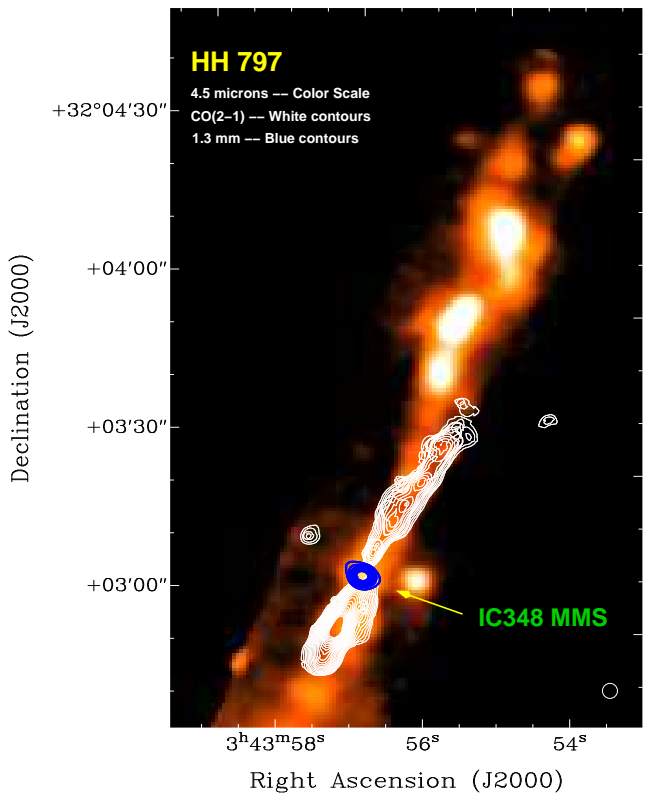}
\caption{\scriptsize SMA CO(2-1) integrated intensity map (moment 0) of  HH 797 (white 
contours) overlaid with a 4.5 $\mu$m Spitzer-IRAC image (color scale), and the 1.4 mm 
continuum emission from the object IC 348-MMS/SMM2 (blue contours). The white contours 
range from 15\% to 80\% of the emission peak in steps of 5\%. The emission peak for the 
CO(2-1) observations is 31 Jy beam$^{-1}$ km s$^{-1}$. The blue contours range from 40\% 
to 90\% of the peak emission in steps of 9\%. The emission peak for the millimeter continuum 
observations is 160 mJy beam$^{-1}$. The synthesized beam of the line and continuum 
observations is shown in the lower right corner.}
\label{fig1}
\end{center}
\end{figure} 

\section{Observations}

The observations were obtained from the SMA archive, and were collected on November 2005, 
when the array was in its compact configuration.  The 21 independent baselines in the compact 
configuration ranged in projected length from 10 to 55 k$\lambda$.  The phase reference center 
for the observations was at $\alpha_{J2000.0}$ = \dechms{03}{43}{57}{29}, $\delta_{J2000.0}$ = 
\decdms{32}{03}{09}{0}. Two frequency bands, centered at 220.538 GHz (Lower Sideband) and 
230.538 GHz (Upper Sideband) were observed simultaneously. The observations were made in 
mosaicing mode using the half-power point spacing between field centers and thus covering the 
entire HH 797 outflow. The primary beam of the SMA at 230 GHz has a FWHM $\sim 50''$. 

The SMA digital correlator was configured in 24 spectral windows (``chunks'') of 104 MHz and
128 channels each. This provides a spectral resolution of 0.812 MHz ($\sim$ 1.1 km s$^{-1}$) 
per channel. The zenith opacity ($\tau_{230 GHz}$), measured with the NRAO tipping 
radiometer located at the nearby Caltech Submillimeter Observatory, fluctuated between 0.12 
and 0.19, indicating reasonable weather conditions during the observations. Observations of 
Uranus provided the absolute scale for the flux density calibration.  The gain calibrators were the 
quasars 3C 111 and 3C 84, while 3C 273 was used for bandpass calibration. The uncertainty in 
the flux scale is estimated to be between 15 and 20$\%$, based on the SMA monitoring of 
quasars.  Further technical descriptions of the SMA and its calibration schemes can be
found in \citet{Hoetal2004}.

The data were calibrated using the IDL superset MIR, originally developed for the Owens Valley 
Radio Observatory \citep[OVRO,][]{Scovilleetal1993} and adapted for the SMA.\footnote{The 
MIR-IDL cookbook by C. Qi can be found at http://cfa-www.harvard.edu/$\sim$cqi/mircook.html.} 
The calibrated data were imaged and analyzed in the standard manner using the MIRIAD and
KARMA \citep{goo96} softwares.  A 1.4 mm continuum image was obtained by averaging line-free 
channels in the upper sideband. We set the ROBUST parameter of the task INVERT to 0 to obtain 
an optimal compromise between resolution and sensitivity. The resulting r.m.s.\ noise for the 
continuum image was about 7 mJy beam$^{-1}$ at an angular resolution of $3\rlap.{''}42$ 
$\times$ $3\rlap.{''}20$ with a P.A. = $-70.8^\circ$. The r.m.s.\ noise in each channel of the 
spectral line data was about 70 mJy beam$^{-1}$ at the same angular resolution.

\begin{figure}[ht]
\begin{center}
\includegraphics[scale=0.35]{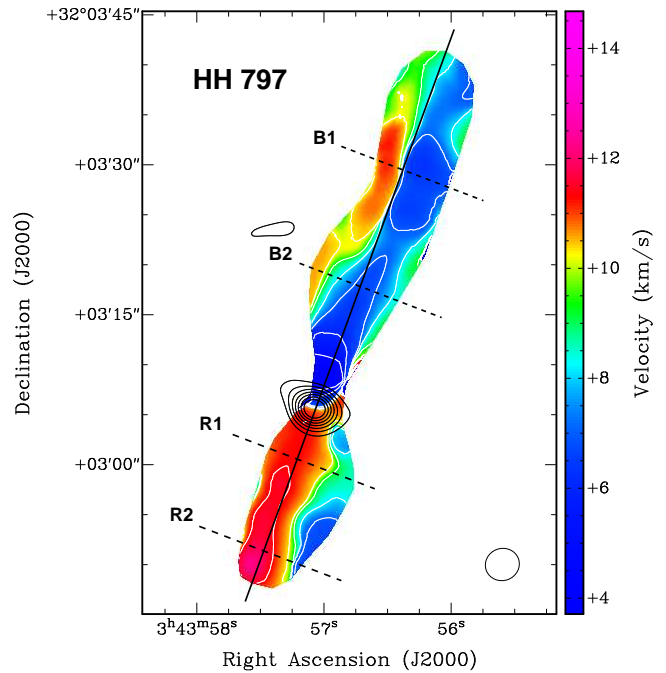}
\caption{\scriptsize SMA CO(2-1) intensity-weighted velocity color map (moment 1) of HH 797 
overlaid in contours with the 1.4 mm continuum emission (black contours) from the circumstellar 
disk associated with IC 348 MMS/SMM2, its exciting source. The black contours range from 
30\% to 90\% of the peak emission, in steps of 10\%. The emission peak for the millimeter 
continuum observations is 160 mJy beam$^{-1}$. The color-scale bars on the right indicate 
the LSR velocities in km s$^{-1}$. The E-W dashed lines mark the positions of the position-velocity 
cuts shown in Figure 3. The N-S continuous black line represents approximately the position of the 
outflow axis. The synthesized beam of the line and continuum observations is shown in the lower 
right corner. }
\label{fig2}
\end{center}
\end{figure}

\begin{figure*}[ht]
\begin{center}
\includegraphics[scale=0.37]{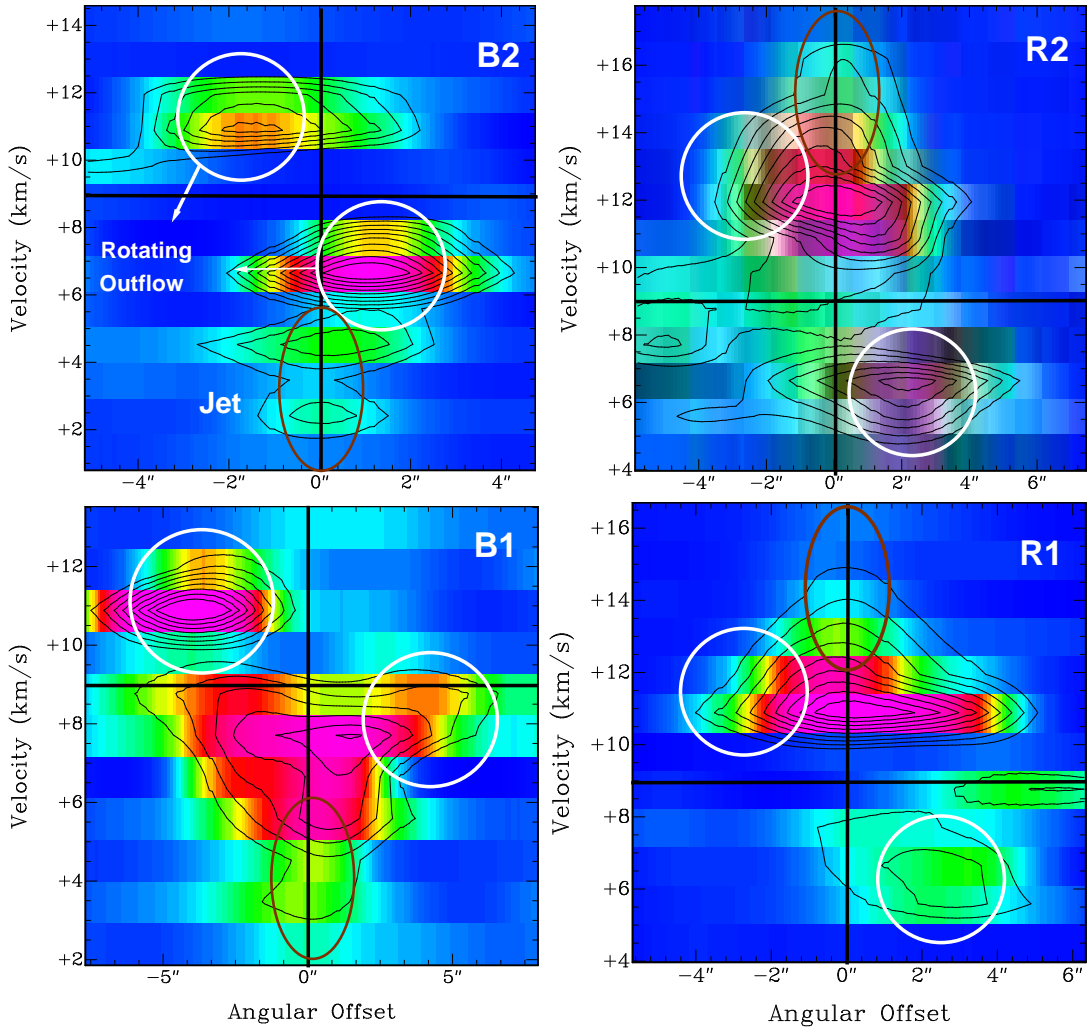}
\caption{\scriptsize Position-velocity diagrams of the transversal cuts marked in Figure 2, across the 
outflow. The black contours range from 30\% to 90\% of the peak emission in steps of 10\%. The black 
lines mark the position of the symmetry axis of the molecular jet and the systemic velocity of the cloud 
(about $+$9 km s$^{-1}$). The velocity and spatial resolutions are approximately 1 km s$^{-1}$ and 
3$''$, respectively. The spatial scale is in arcsec.}
\label{fig3}
\end{center}
\end{figure*} 

\section{Results}

In Figure 1, we show the 1.4 mm continuum map and the integrated intensity $^{12}$CO(2-1) 
line emission from HH 797 overlaid on the Spitzer-IRAC 4.5 $\mu$m image. This IRAC band 
contains several H$_2$ transitions which trace shock-excited material \citep{smi2005}
associated (in the present case) with the supersonic flow in HH 797. This image reveals the 
highly collimated outflow (with an opening angle of about 6$^\circ$ at a position angle of 
$-$30$^\circ$)  already mapped in CO lines by \citet{tafa2006} at a lower spatial resolution. 
From our observations, it is evident that the $^{12}$CO(2-1) emission traces the innermost 
portion ($\sim$ 0.1 pc) of the outflow. The blueshifted radial velocities run from $-$2 to $+$8 
km s$^{-1}$ and the redshifted ones from $+$10 to $+$20 km s$^{-1}$; the systemic LSR 
radial velocity of the ambient molecular cloud is at about $+$ 9 km s$^{-1}$. These velocity 
ranges are in very good agreement with those reported by \citet{tafa2006}. Most of the receding 
radial velocities (redshifted) of the outflow are located toward the south, while the approaching 
radial velocities (blueshifted) are toward the north. We note that the optical and infrared 
components of HH 797 have been extensively discussed by \citet{mac1994}, \citet{eis2003}, 
and \citet{wal2005,wal2006}. 

The outflow emanates from IC 348-MMS/SMM2, a class 0 object \citep{eis2003}. 
Our observations resolve the dust emission from this source, with a deconvolved size 
of (3.0$'' \pm 0.3''$) $\times$ (1.5$'' \pm 0.2''$), 
at P.A.\ $=$  $+50.0^\circ$. This implies a physical size of about 900 $\times$ 450 AU
(assuming a distance of 300 pc) and an orientation almost exactly perpendicular to the outflow 
axis. The source is centered at the $\alpha_{J2000.0}$ = \dechms{03}{43}{57}{073}, 
$\delta_{J2000.0}$ = \decdms{32}{03}{05}{59} and has a flux density AT 1.4 mm
of 220 $\pm$ 15 mJy.  Assuming optically thin isothermal dust emission, a gas-to-dust ratio of 
100, a dust temperature of 40 K, a dust mass opacity $\kappa_{1.4 mm}$ = 1 cm$^2$ g$^{-1}$, 
and an emissivity index $\beta = 1.5$, we estimate a total mass for the source of 0.09 M$_\odot$. 
This combination of properties suggest that the emission seen at 1.4 mm is dominated by the
accretion disk, although a contribution from the inner envelope might also be present. In objects 
where the masses of the disks and central stars have been estimated, they usually are in a ratio 
of order M$_{star}$/M$_{disk}$ $\sim$ 10 \citep{Rod1998, gui1998, sch2006}. This suggests that
the central protostar in IC 348-MMS/SMM2 is roughly of solar mass. 
We searched for a centimeter counterpart using the Very Large Array
archive data from project S9056, taken on 2008 March 13, 14, 18
and 19. We did not detect a source,
setting 4-$\sigma$ upper limits of 0.05 and 0.06 mJy at
4.86 and 8.46 GHz, respectively.

In Figure 2, we present the intensity-weighted velocity map of the $^{12}$CO(2-1) emission, which
confirms that most of the redshifted velocities are found to the south, while the blueshifted velocities 
are to the north. Interestingly, this figure also shows a clear gradient {\em across} the outflow, detected
along the entire length of the flow. This gradient is such that, at a given distance from the protostar, 
the velocity of the gas near the eastern edge of the jet is systematically more red-shifted than that of
the gas on the western edge. 

This asymmetry is confirmed by the position-velocity (PV) diagrams for four cuts across the outflow
(Figure 3), which reveal the spatial structure of the gas across the flow as a function of the radial 
velocities. One can observe that in all cases, a "V"-shape (or, perhaps, more accurately, a triangular 
shape) velocity pattern is apparent. Near the axis of the jet, high and low velocities coexist, extending 
from the systemic velocity up to about $+$ 2 km s$^{-1}$ on the north (blue) side and about  $+$ 16 
km s$^{-1}$ on the south (red) side. In contrast, near the edges of the flow, only comparatively low 
velocities are detected. If the velocity of the gas were parallel to the direction of the outflow, one would 
expect these V shapes to be symmetric with respect to the axis of the flow. In the present case, 
however, there is a clear asymmetry. Consider, for instance, the cuts at positions B1 and B2. Near the 
western edge (at angular offset $\sim +5''$), the bulk gas is at a velocity of about $+$ 8 km s$^{-1}$
(i.e.\ slightly below the systemic velocity, as appropriate for the blue side). But near the eastern
edge (angular offset $\sim -5''$), the gas reaches redshifted velocity of about $+$ 11 km s$^{-1}$.
Similar patterns are seen at positions R1 and R2 and have been reported in two other collimated 
outflows \citep[Orion-S6 and HH 212;][]{lee2008,zap2010} where they have been interpreted as 
evidence for outflow rotation. In the present case, the amplitude of the velocity asymmetry is about
2 km s$^{-1}$. 
We should mention, finally, that the structure of the emission at the systemic velocity (where the 
outflow emission is blended with the extended ambient emission) is poorly reconstructed by our 
interferometric data. The apparent lack of continuity in the PV diagrams at velocity $+$ 9 km 
s$^{-1}$ is a result of this effect.

\section{Discussion}

The most natural interpretation of the velocity asymmetries observed here is jet rotation.  Let us, 
however, consider and discuss alternative explanations. The most obvious alternate interpretation
would be that the flow driven by IC 348-MMS/SMM2 is in fact the superposition of two 
nearly parallel outflows, not spatially resolved by our observations, and powered by a 
central tight binary system.  In a recent study, \citet{mu2008} present a model of such a
situation, and show that the parallel outflows eventually merge, resulting in a persistent 
kink in the final structure. Moreover, the precession induced by the binarity quickly lead to
bending jet trajectories. Although they cannot be entirely ruled out, neither of these effects 
are apparent in our SMA data. In addition, there is currently no independent direct evidence 
supporting the binarity of IC 348-MMS/SMM2.

A second possibility to consider is that of a precessing (non-rotating) jet/outflow system.  This 
possibility would naturally fit with the "wavy" morphology of the outflow at large distances
from the source, as revealed at infrared wavelengths (see Figure 1).  This situation might 
plausibly mimic rotation because the jet would be the superposition of successive events
where gas was ejected in different directions at different times. Such a scenario, however,
does not easily explain the symmetric V-shape morphology seen in the PV diagrams of
HH 797 (which are consistent, instead, with a high velocity jet along the outflow axis). In
addition, precession should produce a point-symmetric situation: if the asymmetries were
east to west along the blue outflow lobe, they should be west to east along the red lobe. In the
present case, however, the asymmetry is consistently east to west along the entire length of the
flow.

Since alternative interpretations fail to reproduce the observed kinematics of the HH 797
flow, we interpret the observed velocity asymmetries as rotation. Considering the situation 
where a jet is launched from a rotating protostellar disk and then accelerated and collimated 
by MHD forces, \citet{Andersonetal2003} provide a formula relating the jet properties 
measured at large distances from the disk to the position (the ``footpoint'') on the disk 
where the jet is anchored.

\begin{eqnarray}
\varpi_{0}\approx0.7\AU\left(\frac{\varpi_{\infty}}
{10\AU}\right)^{2/3}
\left(\frac{v_{\phi,\infty}}{10\kms}\right)^{2/3}\nonumber \\
\left(\frac{v_{p,\infty}}{100\kms}\right)^{-4/3} 
\left(\frac{\stellarmass}{1\solarmass}\right)^{1/3}
\label{eqn:1}
\end{eqnarray}

Here $\varpi_{\infty}$ is the observed radial distance of the jet shell from the flow
axis, $\varpi_{0}$ the radius on the disk from where that shell's material leaves,
v$_{\phi,\infty}$ and v$_{p,\infty}$ are the toroidal and poloidal velocities observed for the shell
at $\varpi_{\infty}$, and $\stellarmass$ the mass of the (proto)star at the center of the disk. In this case, we assume from
observation $\varpi_{\infty}$ $\sim$ 1000 AU (about 4$''$), v$_{\phi,\infty}$ $\sim$ 2 km s$^{-1}$*cos $\alpha$,  v$_{p,\infty}$ 
$\sim$ 10/sin $\alpha$ km s$^{-1}$ and $\stellarmass$ $\sim$ 1 M$_\odot$, with $\alpha$ the unknown angle 
between flow direction and plane of the sky. This results in a footpoint radius of 100*(sin $\alpha)^{4/3}$*(cos $\alpha)^{2/3}$ AU.
Since the apparent poloidal flow velocity of $\sim$ 10 km s$^{-1}$ relative to ambient is relatively low, it seems 
likely that $\alpha$ is small. For $\alpha \le$ 20$^\circ$ the footpoint radius would drop to below 20 AU, a 
reasonable value that fits with the innermost part of the above-mentioned disk dimensions.

The value $\varpi_{\infty}$ = 20 AU is very similar to the one found by \citet{zap2010} for the rotating outflow
Ori-S6 energized by the young class 0 protostar 139-409 and the one estimated by  \citet{Lau2009} for the
the outflow CB 26 powered by a more evolved low-mass T-tauri star. On the other hand, \citet{cho2011,lee2009,lee2008}
have found much smaller values for $\varpi_{\infty}$ in outflows powered for both class 0/1 protostars. Thus, at this point, 
we cannot role out any conclusion about the variation of $\varpi_{\infty}$ with time or some similarities on these values.

\section{Conclusions}

In this paper, we have reported $^{12}$CO(2-1) line observation of HH 797, and discovered velocity asymmetries 
about the flow axis with an amplitude roughly on the order of 2 km s$^{-1}$ over distances of about 1000 AU.  The 
same velocity asymmetries are found on both sides of the outflow, and we interpret them as evidence for jet rotation,
Thus, HH 797 appears to be promising laboratory for future studies of magneto-centrifugal models of jet acceleration. 
\acknowledgments

G. P., L.A.Z, L. L. and L. F. R. acknowledge the financial support from DGAPA, UNAM, and CONACyT, M\'exico. 
L. L. is indebted to the Alexander von Humboldt Stiftung for financial support. We are grateful with the anonymous 
referee whose recomendations helped us to improve the paper. 



{\it Facilities:} \facility{SMA}, \facility{SPITZER (IRAC)}.



\clearpage

\end{document}